\documentclass[12pt,showpacs,preprintnumbers,amsmath,amssymb]{revtex4}

\usepackage[dvips]{epsfig}
\usepackage{latexsym}
\usepackage{psfrag}
\def\setb@se#1{\baselineskip=#1 \normalbaselineskip=#1}

\newcommand{\be}{\begin{equation}}
\newcommand{\ee}{\end{equation}}
\newcommand{\bea}{\begin{eqnarray}}
\newcommand{\eea}{\end{eqnarray}}
\newcommand{\bsub}{\begin{subeqnarray}}
\newcommand{\esub}{\end{subeqnarray}}
\newcommand{\N}{n}
\newcommand{\n}{k}

\newcommand{\W}{v}

\newcommand{\Y}{u}

\newcommand{\ww}{w}
\newcommand{\om}{\sigma}
\newcommand{\thetw}{\theta_{\mbox{{\tiny W}}}}
\newcommand{\mh}{m_{\mbox{{\tiny H}}}}
\newcommand{\mz}{m_{\mbox{{\tiny Z}}}}
\newcommand{\mw}{m_{\mbox{{\tiny W}}}}
\newcommand{\q}{q}

\newcommand{\WW}{{\rm W}}

\newenvironment{subeqnarray}
  {\arraycolsep1pt
    \def\@eqnnum\stepcounter##1{\stepcounter{subequation}{\reset@font\rm
      (\theequation\alph{subequation})}}\eqnarray}%
  {\endeqnarray\stepcounter{equation}}
\makeatother
\begin{document}
\title{Superconducting Electroweak Strings}

\author{Mikhail S. Volkov}
 \affiliation{ {Laboratoire de Math\'{e}matiques et Physique Th\'{e}orique
CNRS-UMR 6083, \\ Universit\'{e} de Tours,
Parc de Grandmont, 37200 Tours, FRANCE}
}
\begin{abstract}
Classical solutions describing strings endowed with an 
electric charge and carrying a constant electromagnetic 
current are constructed  within the bosonic sector of the 
Electroweak Theory. For any given ratio of the Higgs boson mass 
to W boson mass and 
for any Weinberg's angle, these strings comprise a family that  
can be parameterized by values of the  current through their cross section, 
$I_3$, by their electric charge per unit string length, $I_0$,
and by two integers. 
These parameters determine the electromagnetic and Z fluxes, 
as well as the angular momentum and momentum densities of the string.  
For $I_0\to 0$ and $I_3\to 0$ the solutions reduce to  
Z strings, or, for solutions with $I_0=\pm I_3$,   
to the W-dressed Z strings whose existence was discussed some time ago.

\end{abstract}
\pacs{11.15.-q, 11.27.+d, 12.15.-y, 98.80.Cq}
\maketitle

Superconducting strings have been described by Witten within a 
U(1)$\times$U(1) gauge field theory with two complex scalars \cite{Witten}. 
One of these scalars is responsible for the vortex formation, while the other 
one gives rise to a constant current along the string. 
This phenomenon, dubbed `bosonic superconductivity',  has 
found numerous applications, mainly in the cosmological context \cite{review},  
since Witten's model can be viewed as sector of some high energy 
Grand Unification Theory \cite{Witten}. 
One can wonder whether a similar phenomenon could also exist at 
lower energies, as for example in the Electroweak (EW) Theory. 
Indeed, one has also in this case a pair of complex scalar fields,
while the U(1)$\times$U(1) group is contained in the SU(2)$\times$U(1) 
EW gauge group. However, the only EW strings known
up to now have been the Z and W strings 
(see \cite{vacha} for a review), but these carry no current. 

Recently superconducting (SC) strings have been constructed in the semilocal
limit of the EW theory where Weinberg's angle $\thetw=\pi/2$ and the 
Yang-Mills field decouples \cite{SL}. The aim of the present Letter is to 
show that a similar construction can be curried out for 
arbitrary values of $\thetw$.
The gauge field of the resulting 
solutions turns out to be of a quite general
non-Abelian and not of the 
U(1)$\times$U(1) type.
Below we shall show how to obtain them by deforming  
Z strings, which gives solutions physically different 
from Z strings, as is seen by comparing 
the values of the gauge invariant quantities. 
A special case of these new solutions are the W-dressed Z stings 
whose possible existence was discussed some time ago 
\cite{per,Olesen}.

{\bf The EW theory.}
The bosonic sector of the 
theory 
is defined by the action 
$S=\frac{1}{{\rm g}_0^2} \int ({\cal L}_{\rm YM}+{\cal L}_{\rm H}) \, d^4 x$ with 
\bea                             \label{L}
{\cal L}_{\rm YM}=
-\frac{1}{4g^2}\,\WW^a_{\mu\nu}\WW^{a\mu\nu}
-\frac{1}{4g^{\prime 2}}\,{F}_{\mu\nu}{F}^{\mu\nu}, \nonumber \\
{\cal L}_{\rm H}=
(D_\mu\Phi)^\dagger D^\mu\Phi
-\frac{\beta}{8}\left(\Phi^\dagger\Phi-1\right)^2\,,
\eea
where 
$
\WW^a_{\mu\nu}=\partial_\mu\WW^a_\nu
-\partial_\nu \WW^a_\mu
+\epsilon_{abc}\WW^b_\mu\WW^c_\nu
$ 
and 
$
{F}_{\mu\nu}=\partial_\mu{A}_\nu
-\partial_\nu{A}_\mu
$ 
are the SU(2) and U(1) field strengths, respectively, and
$\Phi^{\rm tr}=(\phi_{+},\phi_{-})$ 
is a doublet of complex Higgs fields with 
$
D_\mu\Phi
=\left(\partial_\mu-\frac{i}{2}\,{A}_\mu
-\frac{i}{2}\,\tau^a \WW^a_\mu\right)\Phi
$, where $\tau^a$ are the Pauli matrices. 
Here all fields and spacetime coordinates 
have been rendered dimensionless by rescaling, the 
rescaled gauge couplings are
expressed in terms of the Weinberg angle as 
$g=\cos\thetw$, $g^\prime=\sin\thetw$,
and the dimensionless masses of the Z,W, and Higgs bosons are given,
respectively, by 
${\mz}=1/\sqrt{2}$, 
${\mw}=g \mz$, $\mh=\sqrt{\beta}\,\mz$. 
The mass scale is ${\rm g}_0\Phi_0$ where $\Phi_0$ is the dimensionful 
Higgs field vacuum expectation
value and g$_0$ determines the value of 
the electric charge, $e={\rm g}_0 gg'$.
 The action  is invariant under  
gauge transformations
\bea                            \label{gauge}
\Phi\to {U}\Phi,
~~~~
{\bf W}\to  {U}{\bf W}U^{-1}+2iU\partial_\mu{U}^{-1}dx^\mu,
\eea
where ${\bf W}=(A_\mu+\tau^a W^a_\mu)dx^\mu$ and 
$U\in\,$SU(2)$\times$U(1). 

Varying the action 
gives the field equations, 
\bea
\partial_\mu {F}^{\mu\nu}&=&
g^{\prime 2}\,\Re(i\Phi^\dagger D^\nu\Phi)\,,\nonumber \\
\partial_\mu \WW_a^{\mu\nu}&+&\epsilon_{abc}\WW^b_\sigma\WW^{c\sigma\nu}
=g^{2}\,\Re(i\Phi^\dagger\tau^a D^\nu\Phi)\,,\nonumber \\
D_\mu D^\mu\Phi&=&\frac{\beta}{4}\,(\Phi^\dagger\Phi-1)\Phi.      \label{eqs}
\eea
There exist several ways do define the electromagnetic field ${\cal A}_{\mu\nu}$
(see e.g. \cite{jh}), all of which agree in the Higgs vacuum.
In our case it is convenient to adopt the definition using 
the gauge invariant field tensor \cite{hoft}
\be              \label{hoof}
{\cal W}_{\mu\nu}=\partial_\mu{\cal W}_\nu-\partial_\nu{\cal W}_\mu
-\epsilon_{abc}n^a\partial_\mu n^b\partial_\nu n^c \,,
\ee
where ${\cal W}_\mu=n^a{\rm W}^a_\mu$ and 
$n^a=(\Phi^\dagger\tau^a\Phi)/(\Phi^\dagger\Phi)$. 
The gauge invariant ${\cal A}$ and ${\cal Z}$ fields are then defined as   
${\cal A}_{\mu\nu}=\frac{g}{g^\prime}\, F_{\mu\nu}
-\frac{g^{\prime}}{g}\,{\cal W}_{\mu\nu}$ and
${\cal Z}_{\mu\nu}=F_{\mu\nu}+{\cal W}_{\mu\nu}$.  
The electromagnetic current density is 
$J_\mu=\frac{1}{4\pi}\,\partial^\nu {\cal A}_{\nu\mu}$. 
In the unitary gauge,  one can introduce the potentials
for the ${\cal A,Z}$ fields:
${\cal A}_{\mu}=\frac{g}{g^\prime}\, A_{\mu}
-\frac{g^{\prime}}{g}\,{\cal W}_{\mu}$ and 
${\cal Z}_{\mu}=A_{\mu}+{\cal W}_{\mu}$. 

{\bf Symmetry reduction.} We shall be considering cylindrically symmetric
solutions. Splitting   
the spacetime coordinates as $x^\mu=(x^\alpha,x^k)$, 
where $\alpha=0,3$; $k=1,2$, and introducing the polar coordinates
in the $x^k$ plane, $x^1+x^2=\rho e^{i\varphi}$, we  
assume the system to be invariant under the action of the three
spacetime symmetry generators $K=\{\partial/\partial x^\alpha, 
\partial/\partial\varphi\}$. This implies the conservation of the 
energy,  $E$, momentum, $P$, and angular momentum, $M$, 
expressed (per unit string length) by 
$
\int T^0_\mu K^\mu d^2x
$
for the three above choices of $K$, 
respectively. Here  
$T^\mu_\nu$ is the energy-momentum tensor; $d^2x=dx^1 dx^2$.

Let $\om_\alpha$ be a constant (co)vector in the $x^\alpha$ plane. 
We make the field ansatz  
\bea                           \label{03}
{\bf W}&=&\Y^{+}(\rho)\,\om_\alpha dx^\alpha +
\Y^{-}(\rho)\,d\varphi       \\
&+&
\sum_{a=1,3}{{\tau}^a}\,
[\W^{+}_a(\rho)\,\om_\alpha dx^\alpha + \W^{-}_a(\rho)\, d\varphi],
~~
\phi_{\pm}=f_{\pm}(\rho),~~\nonumber 
\eea
where all functions of $\rho$ are real. 
This ansatz is invariant under complex conjugation and 
it keeps its form also under Lorentz boosts in the $x^\alpha$ plane.
Since the latter preserve the Lorentz norm, 
$
\om^2\equiv -\om_\alpha\om^\alpha=(\om_3)^2-(\om_0)^2,
$
there can be solutions of three different types: magnetic ($\om^2>0$),
electric ($\om^2<0$), and chiral ($\om^2=0$). 
Introducing 
$f=f_{+}+if_{-}$ and $\W_{\pm}=\W^{\pm}_{1}+i\W^{\pm}_{3}$, also 
$X_{\pm}=i\Y^{\pm}\bar{f}+f\W_{\pm}$ and 
$\Lambda=\W_{+}\bar{\W}_{-}-\W_{-}\bar{\W}_{+}$, the field equations obtained by 
inserting \eqref{03} to Eqs.\eqref{eqs} read
\bea            \label{eqs1}
&&\hat{{\cal D}}_{\pm}\Y^{\pm}=\frac{g^{\prime\,2}}{2}\Im\left(fX_{\pm}\right), ~
\hat{{\cal D}}_{\pm}\W_{\pm}=\frac{g^{2}}{2}\bar{f}X_{\pm}
\pm\frac{\Lambda}{2\rho^2}\,\lambda_{\mp}\W_{\mp}\,, \nonumber \\ 
&&\hat{{\cal D}}_{+}f=\sum_{\varsigma=+,-}\frac{\lambda_\varsigma}{4} 
(i\Y^{\varsigma}\bar{X}_{\varsigma}+X_{\varsigma}\bar{\W}_{\varsigma}) 
              +\frac{\beta }{4}(|f|^2-1)f\,, \nonumber \\
&&\Im(\lambda_{+}\W_{+}\bar{\W_{+}}^\prime+\lambda_{-}\,\W_{-}\bar{\W_{-}}^\prime-
g^2f\bar{f}^\prime)=0, 
\eea
where $\hat{{\cal D}}_{\pm}=\frac{d^2}{d\rho^2}\pm\frac{d}{d\rho}$ and 
$\lambda_{+}=\om^2$,
$\lambda_{-}=1/\rho^2$. 
The first order equation in the last line  
is compatible with the other equations 
but constraints the boundary values for their solutions. 

{\bf Boundary conditions.} 
Assuming that 
$u^{-}(0)=k\equiv 2\N-\nu$, 
$\W_1^{-}(0)=0$, $\W_3^{-}(0)=\nu$ with $\N,\nu\in\mathbb{Z}$, 
the fields \eqref{03} can be transformed,
via
gauge transformation with 
$
U=\exp\{-\frac{i}{2}[\eta
+\psi\tau^3]\}\exp\{\frac{i}{2}\gamma\tau^2\}
$ 
where $\psi=V_3^{-}(0)\,\varphi-c_1\om_\alpha x^\alpha$, 
$\eta=k\,\varphi+c_1\om_\alpha x^\alpha$,  $c_1$ is defined in Eq.\eqref{inf},
 to the gauge where they are 
regular at the symmetry axis: 
\bea                           \label{033}
{\bf W}&=&(\Y^{+}-c_1)\,\om_\alpha dx^\alpha +
(\Y^{-}-k)\,d\varphi          
\nonumber \\
&+&
(\tau^1\cos\psi+\tau^2\sin\psi)
[V^{+}_1\om_\alpha dx^\alpha + 
V^{-}_1\, d\varphi]\nonumber \\
&+&\tau^3[(V^{+}_3+c_1)\,\om_\alpha dx^\alpha 
+ (V^{-}_3-V^{-}_3(0))\, d\varphi] \nonumber    \\
&+&\tau^2d\gamma\,,~~~~~~~~
\phi_{\pm}=F_{\pm}e^{-\frac{i}{2}(\eta\pm\psi)}.  
\eea
Here  $V^{\pm}_{1}+iV^{\pm}_{3}=e^{i\gamma}\W_{\pm}$  and 
$F_{+}+iF_{-}=e^{-\frac{i}{2}\gamma}f$, so that $\gamma$ determines the mixing
between the upper and lower components of the Higgs field.  
We shall choose 
either $\gamma=0$ or $e^{\frac{i}{2}\gamma}=-if/|f|$,  
referring to these two choices as gauge IIa and gauge IIb, respectively. 
In both cases $V^{-}_3(0)\in\mathbb{Z}$, 
the latter choice
corresponding to the unitary gauge, $F_{+}=0$. 
The regularity of the Higgs field requires also that 
$F_{\pm}(0)=0$ if $V_3^{-}(0)\pm k\neq 0$. 
The local solution of Eqs.\eqref{eqs1} with such
boundary conditions reads 
\bea                     \label{zero}
\Y&=&a_1+i(\n+a_2\rho^2)+\ldots\,,~
\W_{+}=i+a_{4}\rho^{\nu}+\ldots,~~\nonumber  \\
f&=&a_{3}\rho^{|\N|}+i\q\rho^{|\N-\nu|}+\ldots,~ 
\W_{-}=i\nu+ia_5\rho^2+\ldots~~~ 
\eea
where $a_1,\ldots a_5,q$ are 6 real integration constants. 

 At large $\rho$ we require the fields to approach the 
exact solution of  Eqs.\eqref{eqs1} (written in the gauge \eqref{03}) 
\be           \label{sing}
{\bf W}=(1-\tau^3)[(c_1+Q\ln\rho)\om_\alpha dx^\alpha+c_2 d\varphi],
~~\phi_{+}=1,
\ee
with constant $c_1,c_2,Q$ and with $\phi_{-}=0$, which   
describes fields produced by an electric current distributed along the 
$x^3$ axis.
The corresponding 
asymptotic solution of Eqs.\eqref{eqs1} reads 
\begin{align}                         \label{inf}
&\Y=c_1+Q\ln\rho+ \frac{c_3g^{\prime 2}}{\sqrt{\rho}}e^{-\mz\rho}+i\,[
c_2+ c_4g^{\prime 2}\sqrt{\rho}e^{-\mz\rho}]+\ldots \nonumber \\
&\W_{+}=e^{i c_9}\{\frac{c_5}{\sqrt{\rho}}e^{-m_\om\rho}-i\,[
c_1+Q\ln\rho- \frac{c_3g^{2}}{\sqrt{\rho}}e^{-\mz\rho}
]\}+\ldots
\nonumber \\
&\W_{-}=e^{i c_9}\{c_6\sqrt{\rho}e^{-m_\om\rho}
+i\,[
-c_2+ c_4g^{2}\sqrt{\rho}e^{-\mz\rho}]\}+\ldots \nonumber \\
&f=e^{-\frac{i}{2} c_9}\{1 + \frac{c_7}{\sqrt{\rho}}e^{-\mh\rho}+i\,
\frac{c_8}{\sqrt{\rho}}e^{-m_\om\rho}\}+\ldots
\end{align}
where the dots stand for the higher order and non-linear terms. 
This approaches \eqref{sing}  (modulo the phases) exponentially fast, 
with the rates 
determined by the masses $\mz,\mh$ and $m_\om$, where 
$m_\om^2=\mw^2+\om^2$. 
We thus have a family of local asymptotic solutions parameterized by 
the 10 integration constants  $c_1,\ldots ,c_9$, $Q$.

To construct the global solutions, we
extend the asymptotics \eqref{zero},\eqref{inf} 
and match them in the intermediate region. 
To fulfill the 16 matching conditions for the 8 functions 
and for their first derivatives, we have in our disposal 17
free parameters: 6+10 in Eqs.\eqref{zero},\eqref{inf} and also $\om^2$. 
There is therefore one extra parameter left after matching,
and this will label the global solutions. We choose this parameter to be $\q$
in Eq.\eqref{zero}. 

The above boundary conditions completely determine the    
`worldsheet current' $I_\alpha\equiv \int J_\alpha d^2 x=
-\frac{1}{2gg^\prime}\,\om_\alpha Q$, 
the ${\cal Z}$ flux 
$\Psi_Z=\int {\cal Z}_{\varphi\rho}\, d^2 x=2\pi(k-V_3^{-}(0))$, 
the ${\cal A}$ flux 
$\Psi_A=\int {\cal A}_{\varphi\rho}\, d^2 x=\frac{2\pi}{gg^\prime}
(V_3^{-}(0)-c_2)+
\frac{g}{g^\prime}\Psi_Z$,
and the angular momentum $M=\int T^0_\varphi d^2x=\frac{2\pi}{(gg^\prime)^2}\,
\om_0 Qc_2$.
Here $V_3^{-}(0)$ is computed in the unitary gauge IIb where one goes to 
introduce the potentials to evaluate the fluxes. 
The energy expresses as 
$E=\frac{\pi}{(gg^\prime)^2}(\om_0^2+\om_3^2)Qu^{+}(\infty)+\epsilon$ and the 
momentum is $P=\frac{2\pi}{(gg^\prime)^2}\,\om_0\om_3Q u^{+}(\infty)$,   
where 
\be                               \label{magn}
\epsilon=\int(\frac{\Y_{-}^{\prime 2}}{2\rho^2 g^{\prime 2}}+
\frac{|\W_{-}^{\prime}|^2}{2\rho^2 g^{2}}+\frac{|X_{-}|^2}{4\rho^2}
+|f^\prime|^2+\frac{\beta}{8}(|f|^2-1)^2)d^2x\,.
\ee

{\bf Weakly deformed Z strings.} If $\q=0$ then setting   
$f_{-}=\W^{\pm}_1=0$,  $u^{-}=2g^{\prime 2}(\ww(\rho)-\N)+k$, $\W^{+}_3=-\Y^{+}=1$,
$\W^{-}_3=2g^{2}(\ww(\rho)-\N)+\nu$\,,  Eqs.\eqref{eqs1} reduce to 
\be                  \label{ANO} 
\hat{D}_{+}f_{+}=(\frac{\ww^2}{\rho^2}+\frac{\beta}{4}
(f_{+}^2-1))f_{+},~~~~
\hat{D}_{-}\ww=\frac12\,f_{+}^2\ww,  
\ee 
where
$\N\leftarrow \ww\to 0$ and $0 \leftarrow f_{+}\to 1$ 
as $0\leftarrow \rho\to \infty$.
The solutions are the Z strings \cite{vacha}. In the 
gauge IIa,
\be                         \label{Z}
{\bf W}=2(g^{\prime 2}+g^2\tau^3)(\ww-\N)d\varphi,~~
\phi_{+}=f_{+}e^{-\N\varphi}, 
\ee
and ${\cal Z}_\mu dx^\mu=2(\ww-\N)d\varphi$, ${\cal A}_\mu=0$, 
$\Psi_Z=4\pi\N$.

\begin{figure}[t]
\hbox to\linewidth{\hss%
  \psfrag{x}{$\sin^2\thetw$}
 \psfrag{y}{$\om^2$}
  \psfrag{S21}{$(2,1)$}
  \psfrag{S22}{$(2,2)$}
  \psfrag{S23}{$(2,3)$}
  \psfrag{S24}{$(2,4)$}
  \psfrag{S11}{$(1,1)$}
  \psfrag{S12}{$(1,2)$}
    \resizebox{10cm}{6cm}{\includegraphics{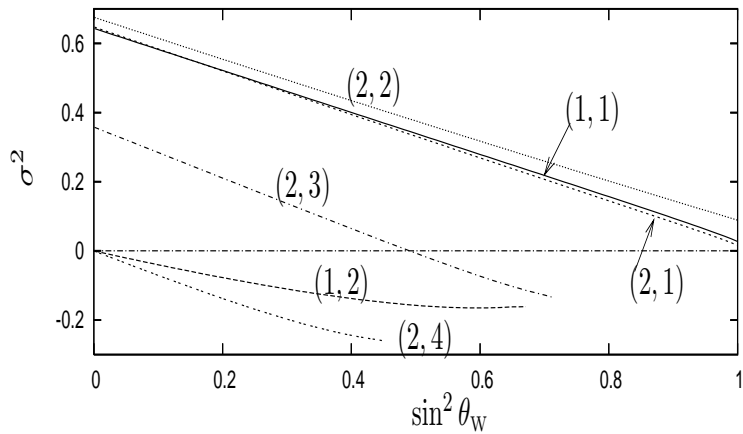}}%
\hss}
\caption{\label{fig1}\small The eigenvalue $\om^2(\beta,\thetw,\N,\nu)$
for  the bound state solutions of Eqs.\eqref{eig} with $\beta=2$, $\N=1,2$.
In the $\om^2<0$ region 
the curves terminate where 
$m_\om=0$.}
\end{figure}

Giving now a small value to $\q$ in 
Eq.\eqref{zero} produces small deformations of the Z strings. 
The linear in $\q$ deformation terms read (always in the gauge IIa)
\be                           \label{0333}
\delta{\bf W}=
(\tau^1\cos\psi+\tau^2\sin\psi)
[\delta\W^{+}_1\om_\alpha dx^\alpha + \delta\W^{-}_1\, d\varphi], 
\ee
also $\delta\phi_{+}=0$
and 
$\delta\phi_{-}=e^{i(\psi-\N\varphi)}\delta f_{-}$. Here
$\delta f_{-}$ and $\delta\W^{\pm}_1$
fulfill equations obtained by linearizing Eqs.\eqref{eqs1} 
around the Z string configuration, which 
can be cast into the eigenvalue problem form  
\be             \label{eig}
\Psi^{\prime\prime}=(\om^2+V[\beta,\thetw,\N,\nu,\rho])\Psi\,,
\ee
 supplemented by the 
linearized version of the constraint equation in \eqref{eqs1}. 
Here $V$ is a $3\times 3$ matrix and $\Psi$ is a 3 component vector. 
Eqs.\eqref{eig} admit 
bound state solutions for which $\Psi\sim e^{-m_\sigma\rho}$ as 
$\rho\to\infty$. 
These describe small deformations of Z strings by a current
$I_\alpha\sim\om_\alpha $, and so the eigenvalue
$\om^2=\om^2(\beta,\thetw,\N,\nu)$
determines the spacetime type of $I_\alpha$.

Given $\beta,\thetw$ and 
$\N\in\mathbb{Z}_{+}$, Eqs.\eqref{eig} admit 
$2\N$ different bound state solutions labeled by $\nu=1,2,\ldots, 2\N$. 
If $\beta>1$ then 
$\N$ of these solutions have $\om^2>0$ for $\thetw\in[0,\frac{\pi}{2}]$.
The eigenvalue $\om^2$
for the other $\N$ solutions is not positive definite, 
and these solutions exist only as long as $m^2_\om=\mw^2+\om^2>0$;  
see Fig.1. Every Z string therefore admits both electric and magnetic 
current deformations. 
Chiral deformations are 
possible if only $\beta,\thetw$
belong to the curves determined by the condition 
$\om^2(\beta,\thetw,\N,\nu)=0$;
see Fig.2. 

Remarkably, these curves 
coincide with those delimiting the Z string stability regions 
\cite{inst}. In fact, the same Z string perturbations as in Eq.\eqref{0333} 
were considered in  \cite{perk,inst},
but choosing $\om_\alpha=(\om_0,0)$ instead of  
$\om_\alpha=(\om_0,\om_3)$. For 
$\om^2=\om_3^2-\om_0^2<0$
these two choices are equivalent,
since one can always boost away the $\om_3$ component in this case.
However, they are physically different for solution of Eqs.\eqref{eig} with
$\om^2>0$. Indeed, if $\om_3=0$ 
then $\om_0$ should be imaginary in this case 
leading to growing in time perturbations,
whereas if $\sigma_3\neq 0$ then  
 $\om_\alpha$ can be a real spacelike vector 
and  the very same solutions will describe the magnetic 
deformations of Z strings. The regions above/below 
a $\om^2(\beta,\thetw,\N,\nu)=0$ curve
therefore correspond to values of $\beta,\thetw$ for which 
the $\nu$-th deformation of the $\N$-th Z string is of the 
electric/magnetic type, 
and also to the values 
for which the $\nu$-th Z string perturbation
mode is stable/unstable.

\begin{figure}[t]
\hbox to\linewidth{\hss%
  \psfrag{x}{$\beta$}
 \psfrag{y}{$\sin^2\thetw$}
  \psfrag{11}{$(1,1)$}
 \psfrag{21}{$(2,1)$}
\psfrag{22}{$(2,2)$}
\psfrag{23}{$(2,3)$}
\psfrag{33}{$(3,3)$}
\psfrag{34}{$(3,4)$}
\psfrag{35}{$(3,5)$}
\psfrag{47}{$(4,7)$}
\psfrag{59}{$(5,9)$}
\psfrag{p}{physical~region}
    \resizebox{10cm}{6cm}{\includegraphics{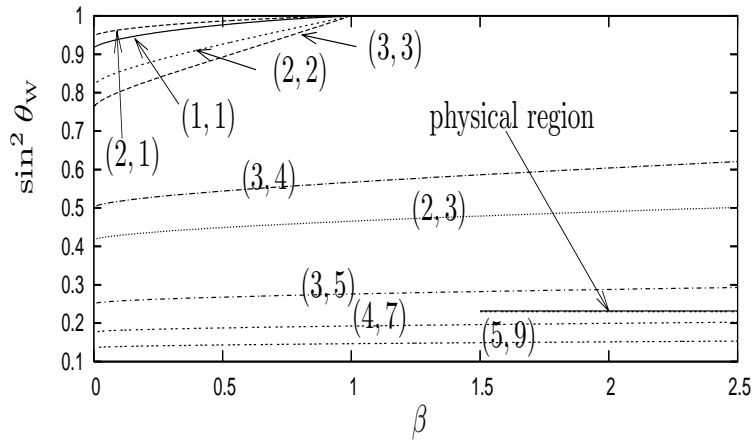}}%
\hss}
\caption{\label{fig2}\small 
The $\om^2(\beta,\thetw,\N,\nu)=0$ curves
for several values of $(\N,\nu)$. Curves with $\nu\leq\N$
are confined to the $\beta\leq 1$ region, while those with $\nu>\N$
extend to higher values of $\beta$.}
\end{figure}  

{\bf SC strings.} Further increasing the value of $\q$ in 
Eq.\eqref{zero} promotes the $2\N$ weak deformations of a given Z string to $2\N$
fully non-linear 
SC  solutions,   
without however changing the sign of $\om^2$, 
so that the above perturbative picture 
remains qualitatively valid. 
The typical solution is shown in Fig.3. These strings carry an electric
current breaking their Lorentz boost invariance, such that each solution 
of Eqs.\eqref{eqs1}
 determines actually a 
whole family 
of strings with fixed $\om^2=-\om_\alpha\om^\alpha$ and with 
$\om_\alpha$'s related to each other by boosts. 
 For given $\beta>0$ and  $\thetw\in[0,\frac{\pi}{2}]$ 
these solutions can be labeled by values of 
$\om_\alpha$, 
by $\N=1,2,\ldots $ and by $\nu=1,\ldots,2\N$.  
The related 
physical parameters of the string are its momentum $P$, angular momentum $M$ 
and charge $I_0$ per unit string length, the ${\cal A,Z}$ 
magnetic fluxes and the current $I_3$ through the string cross section. 
The vector $I_\alpha\sim\om_\alpha$ 
can generically be spacelike or timelike, depending on 
$\nu$.  
For magnetic solutions with $I_\alpha I^\alpha<0$ ($\om^2>0$)
there exists a comoving reference frame 
where $\om_0=I_0=P=M=0$, but their current $I_3$ never vanishes,  
which suggests the term
`superconductivity'. 
For electric solutions with $I_\alpha I^\alpha>0$ there exists a 
reference frame where $\om_3=I_3=P=0$, but the charge $I_0$ never vanishes.  
\begin{figure}[h]
\hbox to\linewidth{\hss%
  \psfrag{x}{$\ln(1+\rho)$}
  \psfrag{f}{W$^2_\rho$}
  \psfrag{p}{$F_{-}$}
 \psfrag{z}{$u^{-}$}
 \psfrag{w1}{$V_1^{-}$}
 \psfrag{w3}{$V_3^{-}$} 
  \psfrag{o1}{$V_1^{+}$}
  \psfrag{y}{$\frac12(u^{+}-V_3^{+})$} 
 \psfrag{o3}{$u^{+}$}
  \resizebox{10cm}{6cm}{\includegraphics{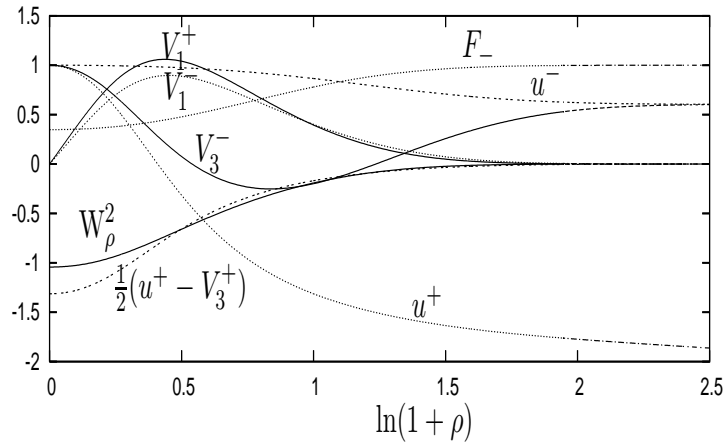}}%
\hss}
\caption{\label{fig3}\small 
Magnetic SC solution with 
$\beta=2$, $g^{\prime 2}=0.23$, $\N=\nu=1$ and  
$\q=0.348$ shown in the unitary gauge IIb.
}
\end{figure}

The rest frame values of the charge $I_0$, current $I_3$, angular 
momentum $M$, and also of $\om^2$, $\Psi_Z$, $\Psi_A$ 
for the $\nu=1,2$ SC solutions with $\N=1$,
$\beta=2$, $\sin^2\thetw=0.23$ are shown in Fig.4 as functions of $\q$. 
We observe that, although for the electric solution the fluxes 
$\Psi_Z$, $\Psi_A$ reduce to their Z string values as $\q\to 0$, 
this does not happen for the magnetic solution, although  
all SC solutions converge to the same Z string limit 
as $\q\to 0$ (in the gauge IIa). 
To understand this we notice that in the   
unitary gauge IIb used to compute the fluxes 
one has 
$V_3^{-}(0)=-\nu$ for $\q=0$ and $V_3^{-}(0)=\nu$
for $\q\neq 0$ ($\nu<2\N$).   
SC strings are thus indeed `topologically different' and converge to 
Z strings only pointwise in this gauge.
Computing the fluxes 
gives $\Psi_Z=4\pi\N$ for Z strings, while 
for the SC strings 
$\Psi_Z(\nu)=4\pi (\N-\nu)$ for $\nu=1,\ldots ,2\N-1$ and 
$\Psi_Z(2\N)=4\pi(\N-\nu\q^2/(\q^2+a_3^2))$.

Since the SC strings are coupled to the 
electromagnetic field, their energy per unit length (also $P$)
is divergent: $E$ contains the term  
$Q(\om_0^2+\om_3^2)\Y^{+}(\infty)$ where $\Y^{+}(\rho)\sim\ln\rho$
for $\rho\to\infty$. 
This type of divergence arises, however, even for 
ordinary electrical wires, if they are infinitely long, 
whereas objects made of 
finite pieces of strings, such as closed current loops, 
will have finite energy. 
$E$  becomes finite 
if $g=0$ or $g^\prime=0$, since the long range massless fields then 
decouple and $Q=0$. 
For $g=0$ the solutions reduce to the twisted SC semilocal strings 
studied in \cite{SL}. 

\begin{figure}[h]
\hbox to\linewidth{\hss%
  \psfrag{x}{{$\q$}}
  \psfrag{SIG}{$\om^2$}
  \psfrag{PsiA}{$\Psi_A/4\pi$}
 \psfrag{CUR}{$|I_3|$}
 \psfrag{PsiZ}{$\Psi_Z$}
 \psfrag{nu}{$\nu=1$} 
 \psfrag{2SIG}{$4\om^2$}
  \psfrag{2PsiA}{$\Psi_A$}
 \psfrag{2CUR}{$4|I_0|$}
 \psfrag{2PsiZ}{$\Psi_Z/4\pi$}
 \psfrag{2nu}{$\nu=2$} 
\psfrag{2M}{$M$} 
  \resizebox{8cm}{6cm}{\includegraphics{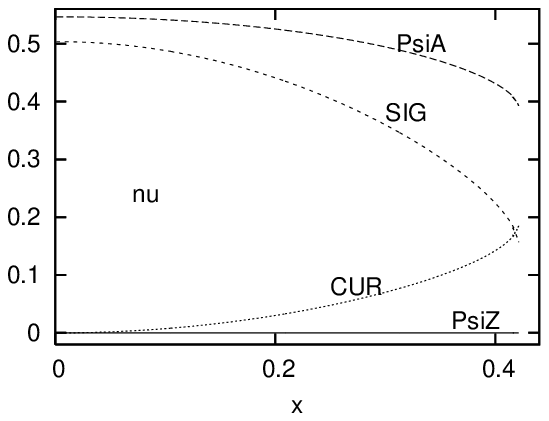}}%
  \resizebox{8cm}{6cm}{\includegraphics{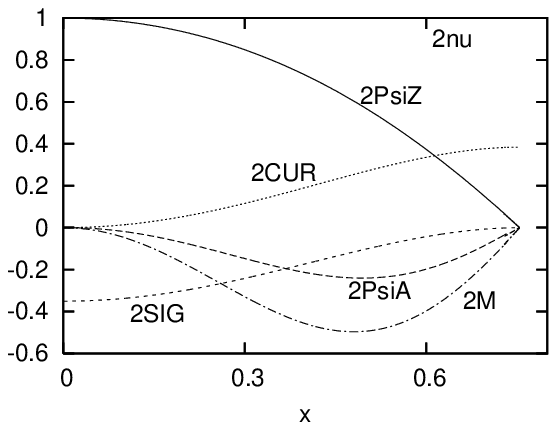}}%
\hss}
\caption{\label{fig4}\small 
The 
parameters of the SC solutions with $\N=1$,
$\beta=2$, $\sin^2\thetw=0.23$, and $\nu=1,2$.    
}
\end{figure}

{\bf W-dressed Z strings.} $E$ is also finite 
for special chiral solutions with 
$\om_\alpha=I_\alpha=0$.
These can presumably be viewed as 
superpositions of SC strings with currents
flowing in the opposite directions and canceling each other.
Chiral solutions are not
generic, given $\q\neq 0$ they exist only for  values of $\beta,\thetw$ 
belonging to the curves determined by the condition 
$\om^2(\beta,\thetw,\N,\nu,\q)=0$.
For $\q\to 0$ these curves reduce to those shown in Fig.2, 
while for $\q\neq 0$ they remain qualitatively the same but shift  
{\it upwards}.
As a result, given a point $\beta,\thetw$ in an 
(in fact small) upper vicinity of a chosen 
curve in Fig.2, one can adjust the value of $\q$ such that the shifted 
curve will pass
through this point. This fine tuning determines 
the values of $\beta,\thetw,\q$ giving rise 
to a solution of Eqs.\eqref{eqs1} with $\om^2=0$. Setting then 
$\om_\alpha=0$, this solution describes a static, 
purely magnetic curentless string 
which is not a gauge copy of a Z string \cite{perk} 
since it has $\Psi_A\neq 0$. The profiles of such solutions are 
shown in Fig.5, where 
$\Y^{+},V^{+}_a$ are no longer needed since they are multiplied by 
 $\om_\alpha=0$ in the ansatz. 
Such solutions apparently correspond to the 
W-dressed Z strings discussed some time ago 
\cite{per,Olesen}. These were supposed to be  
strings with a W-condensate at the core, with lower energy than 
Z strings, and hopefully with better stability. 
They were looked for numerically \cite{perk},  
but with negative results, presumably  
because their parameters were not fine tuned. 
We find that for $\N=1$ these dressed solutions  
are indeed energetically 
favored as compared to Z strings with the same $\beta,\thetw$. However, 
the parameters  $\beta,\thetw$ lie then in the  
unphysical region (see Fig.2). 
The dressed solutions also exist in the physical region, for   
$g^{\prime 2}=0.23$ and $1.5\leq\beta\leq 3.5$, 
 but only for higher values of 
$\N,\nu$ starting from $\N=4$, $\nu=7$ (see Fig.2), in which case we find them  
to be slightly more heavy than Z strings; see Fig.5.

\begin{figure}[h]
\hbox to\linewidth{\hss%
  \psfrag{x}{{$\ln(1+\rho)$}}
  \psfrag{f}{W$^2_\rho$}
  \psfrag{p}{$F_{-}$}
 \psfrag{z}{$u^{-}$}
 \psfrag{w1}{$V_1^{-}$}
 \psfrag{w3}{$V_3^{-}$} 
  \resizebox{8cm}{6cm}{\includegraphics{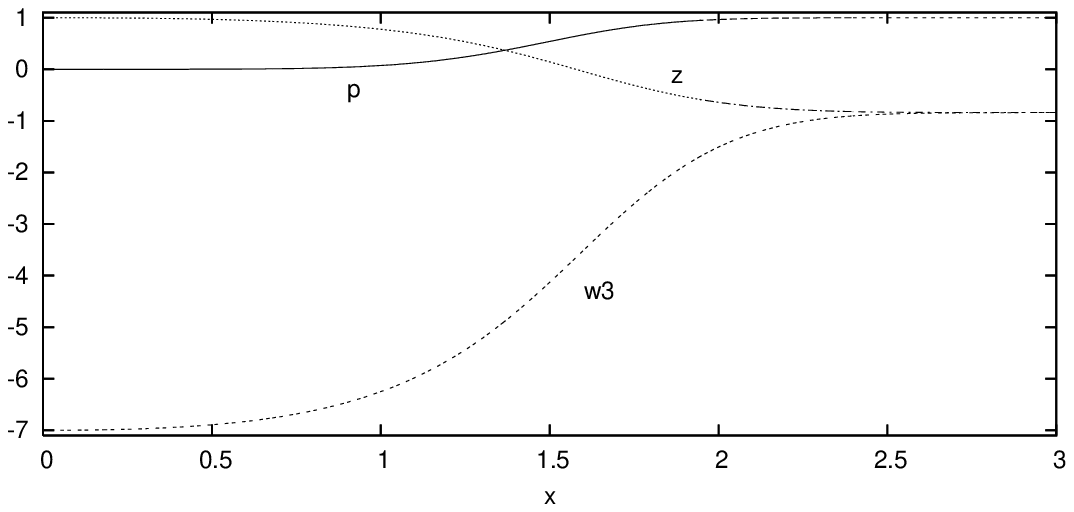}}%
  \resizebox{8cm}{6cm}{\includegraphics{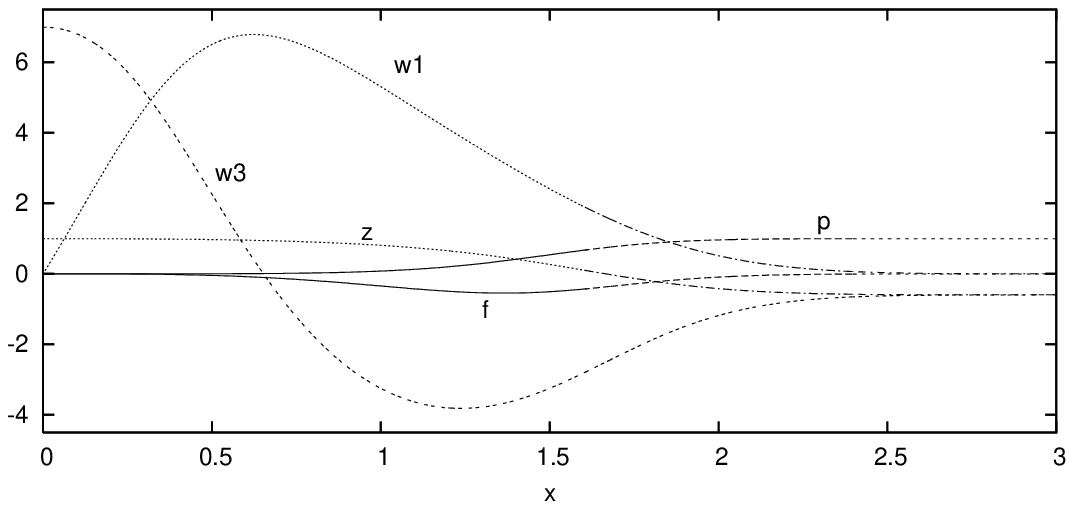}}%
\hss}
\caption{\label{fig5}\small 
Bare Z string (left) and W-dressed Z string (right) solutions with 
$\beta=2$, $\sin^2\thetw=0.23$, $\N=4$, $\nu=7$. They have $E=5.03$
and $E=5.04$, respectively.    
}
\end{figure}

Summarizing, we have sketched the construction of a new class of 
non-topological string solutions in 
Standard Model. Comprising a large family, 
they extend considerably the knowledge of the non-perturbative 
sector of the theory, while their superconductivity may lead to 
interesting physical effects similar to those discussed in 
\cite{review}.  Their conserved current may have 
a stabilizing effect on them, although additional 
investigations are necessary to clarify the issue of stability of these
solutions.

It is a pleasure to acknowledge 
discussions with 
Ana Achucarro, Mark Hindmarsh, 
Mikhail Shaposhnikov, Paul Shellard, and Tanmay Vachaspati.   
This work was supported in part by the ANR under grant NT05-1$_{-}$42856.

\end{document}